\documentclass[conference]{ieeeconf}

\IEEEoverridecommandlockouts                              

\usepackage{xr}

\newcommand{\omitted}[1]{}

\title{
Performance-Aware Self-Configurable Multi-Agent Networks: A Distributed Submodular Approach for Simultaneous Coordination and Network Design
}
\author{Zirui Xu, Vasileios Tzoumas$^\dagger$
	\thanks{
 $^\dagger$Department of Aerospace Engineering, University of Michigan, Ann Arbor, MI 48109 USA;  {\tt\footnotesize \{ziruixu,vtzoumas\}@umich.edu}} 
    \thanks{This work was partially supported by NSF CAREER No. 2337412.}
}

\usepackage{cite}

\usepackage{comment}
\usepackage{siunitx}
\usepackage{relsize}
\usepackage{ifthen}
\usepackage[colorinlistoftodos]{todonotes}

\usepackage[vlined,ruled,linesnumbered]{algorithm2e}
\usepackage{graphics} 
\usepackage{rotating}
\usepackage{color}
\usepackage{enumerate}
\usepackage[T1]{fontenc}
\usepackage{psfrag}
\usepackage{epsfig} 
\usepackage{booktabs}
\usepackage{graphicx,url}
\usepackage{multirow}
\usepackage{array}
\usepackage{latexsym}
\usepackage{amsfonts}
\usepackage{amsmath}
\usepackage{amssymb}
\usepackage{xstring}
\usepackage{algorithmic}
\usepackage{multirow}
\usepackage{xcolor}
\usepackage{prettyref}
\usepackage{flexisym}
\usepackage{bigdelim}
\usepackage{breqn} 
\usepackage{listings}
\let\labelindent\relax
\usepackage{xspace}
\usepackage{bm}
\graphicspath{{./figures/}}
\usepackage{tikz}
\usetikzlibrary{matrix,calc}
\usepackage{lipsum}
\usepackage{mdwlist}

\let\itemize\undefined
\makecompactlist{itemize}{stditemize}
\usepackage{enumitem}
\usepackage{caption}
\usepackage{amsthm}
\usepackage{mathtools}

\makeatletter
\let\NAT@parse\undefined
\makeatother
\usepackage{hyperref}
\usepackage{cleveref}

\hypersetup{%
  colorlinks=true,
  linkcolor=blue,
  filecolor=magenta,      
  urlcolor=black,
  citecolor=red,
  linkbordercolor={0 0 1}
}

\newtheorem{theorem}{Theorem}
\newtheorem{problem}{Problem}

\newtheorem{corollary}{Corollary}

\newtheorem{lemma}{Lemma}

\newtheorem{definition}{Definition}
\newtheorem{proposition}{Proposition}

\newtheorem{remark}{Remark}

\newcommand{\bdmath}{\begin{dmath}}
\newcommand{\edmath}{\end{dmath}}
\newcommand{\beq}{\begin{equation}}
\newcommand{\eeq}{\end{equation}}
\newcommand{\bdm}{\begin{displaymath}}
\newcommand{\edm}{\end{displaymath}}
\newcommand{\bea}{\begin{eqnarray}}
\newcommand{\eea}{\end{eqnarray}}
\newcommand{\beal}{\beq \begin{array}{lll}}
\newcommand{\eeal}{\end{array} \eeq}
\newcommand{\beas}{\begin{eqnarray*}}
\newcommand{\eeas}{\end{eqnarray*}}
\newcommand{\ba}{\begin{array}}
\newcommand{\ea}{\end{array}}
\newcommand{\bit}{\begin{itemize}}
\newcommand{\eit}{\end{itemize}}
\newcommand{\ben}{\begin{enumerate}}
\newcommand{\een}{\end{enumerate}}

\newcommand{\calA}{{\cal A}}
\newcommand{\calB}{{\cal B}}
\newcommand{\calC}{{\cal C}}

\newcommand{\calE}{{\cal E}}

\newcommand{\calG}{{\cal G}}

\newcommand{\calJ}{{\cal J}}

\newcommand{\calM}{{\cal M}}
\newcommand{\calN}{{\cal N}}

\newcommand{\calV}{{\cal V}}

\definecolor{myblue}{RGB}{65 105 225}

\newcommand{\hide}[1]{}

\newcommand{\hiddenText}{{\color{gray} hidden text.}}
\newcommand{\hideWithText}[1]{\hiddenText}

\newcommand{\opt}{^{\star}}

\newcommand{\scenario}[1]{{\fontsize{8.5}{8.5}\selectfont\sf #1}\xspace}

\newcommand{\ie}{\emph{i.e.},\xspace}

\newcommand{\myin}{\, \in \,}

\newcommand{\blue}[1]{{\color{blue}#1}}

\newcommand{\dfs}{\scenario{DFS-SG}}

\newcommand{\myParagraph}[1]{{\bf #1.}\xspace}

\renewcommand{\opt}{\scenario{OPT}}

\newcommand{\curv}{\kappa}

\newcommand{\alg}{\scenario{Anaconda}}
\newcommand{\actionsel}{\scenario{ActionCoordination}}
\newcommand{\neighborsel}{\scenario{NeighborSelection}}

\newcommand{\elem}{v}

\newcommand{\distfsf}{p}

\newcommand{\solopt}{\calA^{\opt}}

\newcommand{\AReg}{\operatorname{A-Reg}_{T}}
\newcommand{\NReg}{\operatorname{N-Reg}_{\{a_t\}_{t\in [T]}}}

\newcommand{\smi}[2]{I_{f,t}({#1};\,{#2})}

\PassOptionsToPackage{end}{algorithmic}

\renewcommand{\baselinestretch}{0.98}

\begin{document}

\maketitle

\thispagestyle{empty}
\pagestyle{empty}

\begin{abstract}
We introduce the first, to our knowledge, rigorous approach that enables multi-agent networks to self-configure their communication topology to balance the trade-off between scalability and optimality during multi-agent planning.  
We are motivated by the future of ubiquitous collaborative autonomy where numerous distributed agents will be coordinating via agent-to-agent communication to execute complex tasks such as traffic monitoring, event detection, and environmental exploration. 
But the explosion of information in such large-scale networks currently curtails their deployment due to impractical decision times induced by the computational and communication requirements of the existing near-optimal coordination algorithms.
To overcome this challenge, we present the {AlterNAting COordination and Network-Design Algorithm} (\alg), a scalable algorithm that also enjoys near-optimality guarantees. 
Subject to the agents' bandwidth constraints, \alg enables the agents to optimize their local communication neighborhoods such that the action-coordination approximation performance of the network is maximized. Compared to the state of the art, \alg is an anytime self-configurable algorithm that quantifies its suboptimality guarantee for any type of network, from fully disconnected to fully centralized, and 
that, for sparse networks, is one order faster in terms of decision speed.
To develop the algorithm, we quantify the suboptimality cost due to decentralization, \ie due to communication-minimal distributed coordination.  We also employ tools inspired by the literature on multi-armed bandits and submodular maximization subject to cardinality constraints.  We demonstrate \alg in simulated scenarios of area monitoring and compare it with a state-of-the-art algorithm. 
\end{abstract}

\vspace{-3.3mm}
\section{Introduction}\label{sec:Intro}

In the future, distributed teams of agents will be coordinating via agent-to-agent communication to execute tasks such as {target tracking}~\cite{xu2023bandit}, {environmental mapping}~\cite{atanasov2015decentralized}, and {area monitoring}~\cite{corah2018distributed}. Such multi-agent tasks are modeled in the robotics, control, and machine learning literature via maximization problems of the form
\begin{equation}\label{eq:intro}
    \vspace{-.5mm}
	\max_{a_{i}\,\in\,\mathcal{V}_i,\,  \forall\, i\,\in\, \calN}\
	f(\,\{a_{i}\}_{i\myin \calN}\,), \vspace{-.5mm}
\end{equation}
where $\calN$ is the set of agents, $a_{i}$ is agent $i$'s action, $\calV_i$ is agent $i$'s set of available actions, and $f\colon 2^{\prod_{i \in \calN}\calV_i}\mapsto\mathbb{R}$ is the objective function that captures the task utility~\cite{krause2008near,singh2009efficient,tokekar2014multi,atanasov2015decentralized,gharesifard2017distributed,marden2017role,grimsman2019impact,corah2018distributed,schlotfeldt2021resilient,xu2022resource,du2022jacobi,rezazadeh2023distributed,robey2021optimal}. Particularly, in information gathering tasks, $f$ is often \textit{submodular}~\cite{fisher1978analysis}: submodularity is a diminishing returns property, and it emanates due to the possible information overlap among the information gathered by the agents~\cite{krause2008near}. For example, in target monitoring with multiple cameras at fixed locations, $\calN$ is the set of cameras,  $\calV_i$ is the available directions the camera can point at, and $f$ is the number of targets covered by the collective field of view of the cameras. 

But solving the problem in~\cref{eq:intro} in real-time is challenging since it is NP-hard~\cite{Feige:1998:TLN:285055.285059}. Although polynomial-time

\begin{figure}[t]
    \captionsetup{font=footnotesize}
    \centering
\hspace{0.4mm}\includegraphics[width=.99\columnwidth]{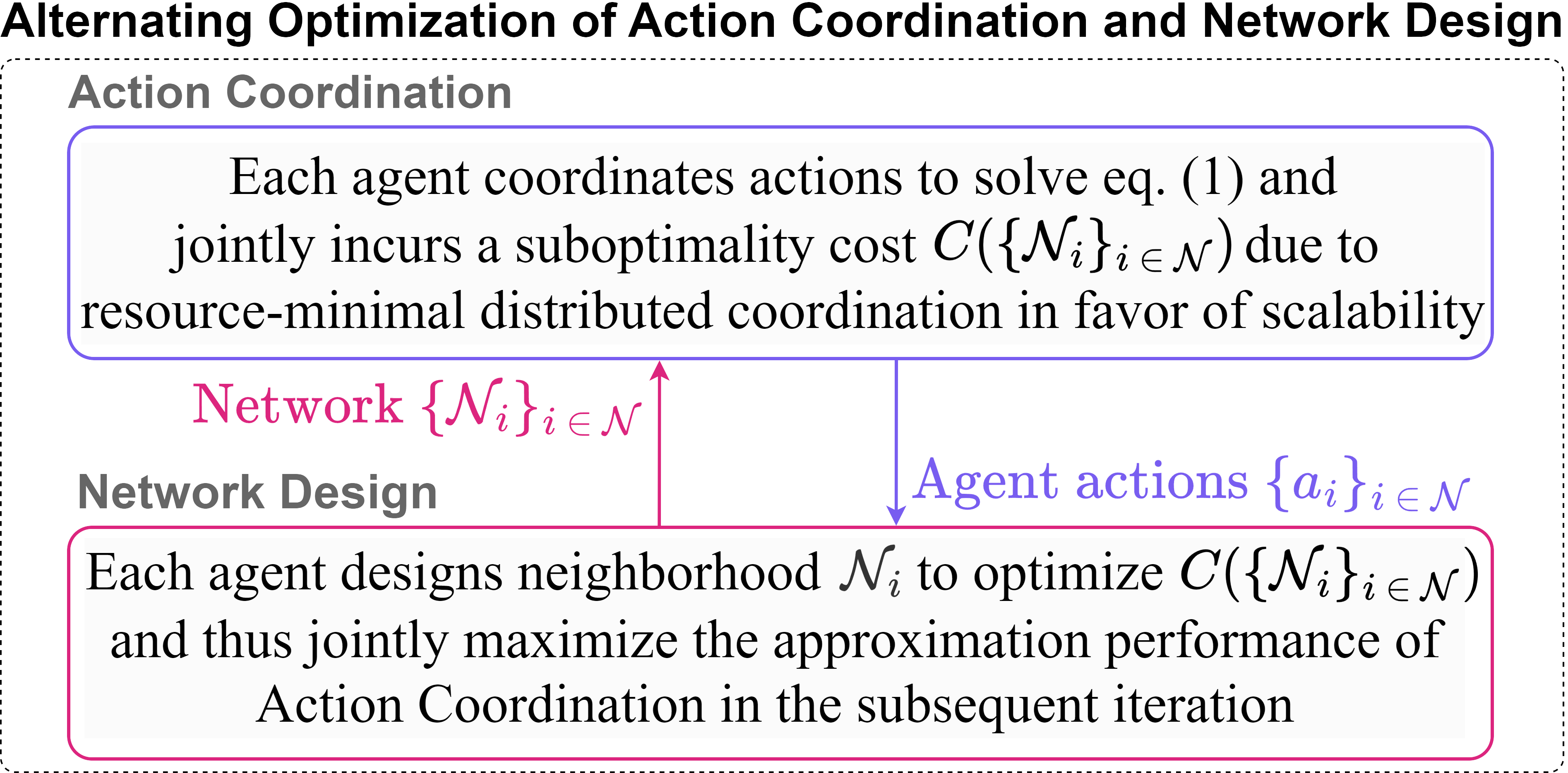}
    \vspace{-5mm}
    \caption{\textbf{Overview of AlterNAting COordination and Network-Design Algorithm ({\fontsize{7}{7}\selectfont\sf Anaconda})}. Starting from an unspecified communication network, and subject to the agents' communication bandwidth and connectivity constraints, {\fontsize{7}{7}\selectfont\sf Anaconda} enables the agents to optimize their local communication neighborhoods such that the action-coordination approximation performance of the whole network is maximized. To this end, {\fontsize{7}{7}\selectfont\sf Anaconda} employs two subroutines, {\fontsize{7}{7}\selectfont\sf ActionCoordination} and {\fontsize{7}{7}\selectfont\sf NeighborSelection}, that alternate optimization.  In more detail,  given the selected neighborhoods $\{\calN_i\}_{i\in\calN}$ by {\fontsize{7}{7}\selectfont\sf NeighborSelection}, {\fontsize{7}{7}\selectfont\sf ActionCoordination} instructs the agents to select actions to jointly maximize \cref{eq:intro}. But {\fontsize{7}{7}\selectfont\sf ActionCoordination} incurs a suboptimality cost $C(\{\calN_i\}_{i\in\calN})$ due to requiring the agents to coordinate exchanging local information only, prohibiting also multi-hop communication, in favor of decision speed.  For this reason, given the agents' bandwidth and connectivity constraints, and the previously selected actions by {\fontsize{7}{7}\selectfont\sf ActionCoordination},  {\fontsize{7}{7}\selectfont\sf NeighborSelection} instructs each agent~$i$  to design its neighborhood $\calN_i$ to optimize $C(\{\calN_i\}_{i\in\calN})$ and, thus, maximize the approximation performance of {\fontsize{7}{7}\selectfont\sf ActionCoordination} in the subsequent iteration. 
    }\label{fig:overview}
   \vspace{-7mm}
\end{figure}

\noindent algorithms exist that achieve near-optimal solutions for the problem in \cref{eq:intro}, in the presence of real-world communication delays~\cite{wu2021comprehensive}, these algorithms often require high times to terminate ---communication delays are caused by the finite speed of real-world communication channels.  The reason is that, for an increasing number of agents, the current algorithms require a combination of high number of communication rounds and large inter-agent message lengths, which collectively increase the total delay.  

For example, the algorithm in~\cite{robey2021optimal}, although it achieves the approximation bound $1-1/e$ for~\cref{eq:intro}, which is the best possible~\cite{calinescu2011maximizing}, can require tenths of minutes to terminate even for $10$ agents~\cite{xu2022resource}.  This is due to algorithm in~\cite{robey2021optimal} requiring near-cubic communication rounds in the number of agents, and inter-agent messages that carry information about all agents, instead of only local information.  Similarly, the Sequential Greedy algorithm~\cite{fisher1978analysis}, which is the gold standard in robotics, control, and machine learning~\cite{krause2008near,singh2009efficient,tokekar2014multi,atanasov2015decentralized,gharesifard2017distributed,marden2017role,grimsman2019impact,corah2018distributed,schlotfeldt2021resilient,xu2022resource,du2022jacobi,rezazadeh2023distributed,robey2021optimal}, although it sacrifices some approximation performance to enable faster decision speed ---achieving the bound $1/2$ instead of the bound $1-1/e$--- still requires (i) inter-agent messages that carry information about all the agents and, in the worst case, (ii) a quadratic number of communication rounds over directed networks~\cite[Proposition~2]{konda2022execution}, resulting in a communication complexity that is cubic in the number of agents, as we will discuss later in \Cref{rem:vsSG}. 

In a similar vein, the Resource-Aware distributed Greedy (\scenario{RAG}) algorithm~\cite{xu2022resource} aims to sacrifice even further approximation performance in favor of more scalability, by requiring the agents to receive information only from and about neighbors such that each message contains information only about the neighbor that sends it.  In more detail, the messages are received directly from each neighbor, via multi-channel communication, assuming a pre-defined directed communication network that respects all agents' communication bandwidths.  That way, \scenario{RAG} enables (i) parallel decision-making, instead of only sequential that the Sequential Greedy requires, reducing the total number of communication rounds it (\scenario{RAG}) requires, and (ii) inter-agent messages that contain information about one agent only, instead of multiple agents that the Sequential Greedy requires, thus making the communication of such shorter messages faster in the presence of finite communication speeds. Due to \scenario{RAG}'s communication-minimal protocol, \scenario{RAG} has an approximation performance that is the same as the Sequential Greedy when all agents are neighbors with all other ---fully centralized coordination--- but, when agents coordinate with a few others only, \scenario{RAG} suffers a suboptimality cost as a function of the network topology.  Therefore, the following research question arises: 

\vspace{0.5mm}
\textit{Subject to the agents' bandwidth constraints, how to enable each agent to optimize its coordination neighborhood such that the suboptimality cost due to decentralization is minimized, that is, the action-coordination performance of the multi-agent network is maximized?}
\vspace{0.5mm}

To our knowledge, no current work provides distributed algorithms that design the network topology to rigorously optimize the coordination's approximation performance. Heuristic methods are proposed for network optimization yet without being rigorously tied to optimizing the coordination performance~\cite{liu2020who2com,niu2021multi}. Although~\cite{xu2022resource} quantifies the suboptimality cost due to decentralization, it cannot be leveraged to enable the agents to optimize their neighborhoods since it requires the agents to have oracle access to the actions of non-neighbors, which is impossible in practice in favor of scalability.  Works also provide fundamental limits in using the Sequential Greedy for distributed submodular optimization where the agents can select actions ignoring the actions of some of the previous agents~\cite{gharesifard2017distributed,marden2017role,grimsman2019impact}.  Particularly,~\cite{gharesifard2017distributed,grimsman2019impact} assume a Directed Acyclic Graph (DAG) information-passing communication topology.
In this context, these works characterize Sequential Greedy's worst-case performance via graph-theoretic properties of the network. For example,~\cite{grimsman2019impact} characterizes the worst-case approximation bound by considering the worst-case over all submodular functions, and proves a bound that scales 
inversely proportional to the independence number of the information graph.  Intuitively, the bound scales inversely proportionally to the maximum number of groups of agents that can plan independently.  
In contrast, in this work, we provide algorithms that allow for the distributed co-design of the network topology based on the submodular function $f$ at hand and the agents' bandwidth constraints.  Even if all agents plan independently (fully decentralized network), the guaranteed bound can still be $1/2$, depending on $f$'s curvature (\Cref{th:main}), instead of scaling inversely proportional to the number of agents.

\myParagraph{Contributions}
We provide the first, to our knowledge, rigorous approach that enables multi-agent networks to self-configure their communication topology to balance the trade-off between scalability and optimality during multi-agent coordination.  To this end,  we initiate the study of the problem \textit{Distributed Simultaneous Coordination and Network Design} (\Cref{sec:problem}), and present a scalable online algorithm with near-optimality guarantees (\Cref{sec:algorithm,sec:tracking-regret,sec:resource-guarantees}). 

Subject to the agents' bandwidth and communication constraints, \textit{the algorithm enables the agents to optimize their local communication neighborhoods such that the action-coordination approximation performance of the whole network is maximized}. The optimization occurs over multiple rounds of local information exchange, where information relay via multi-hop communication is prohibited to curtail the explosion of information exchange, and, thus, to keep low delays due to limited communication speeds. 
We overview the algorithm in more detail in Fig.~\ref{fig:overview}. 

To enable the algorithm, we introduce the first, to our knowledge, quantification of the suboptimality cost during distributed coordination as a function of each agent's neighborhood (\Cref{sec:tracking-regret}).  To this end, we capture the action overlap through $f$ between each agent and its neighbors via a mutual-information-like quantity that we term \textit{Mutual Information between an Agent and its Neighbors} (\Cref{def:MI}). 

The algorithm has the following properties:

\paragraph{Anytime Self-Configuration} The algorithm
enables each agent to select actions and neighbors based on local information only. Therefore, the algorithm
enables the multi-agent system to fluidly adapt to new near-optimal actions and communication topology whenever either new agents are included to or existing agents are removed from the network. 
 
\paragraph{Decision Speed} For sparse networks, where the size of each agent's neighborhood is \underline{not} proportional to the size of the whole network, the algorithm is an order faster than the state-of-the-art algorithms when accounting for the impact that the message length has to communication delays (\Cref{sec:resource-guarantees}).
The result holds true when the communication cost to the decision speed is at least as high as the computational cost. 
Particularly, we quantify the algorithm's decision speed in terms of the time needed to perform function evaluations and to communicate with finite communication speeds. 
	
\paragraph{Approximation Performance} 
The algorithm enjoys a suboptimality bound against an optimal fully centralized algorithm for \cref{eq:intro} (\Cref{sec:tracking-regret}).  Particularly, the bound:
\begin{itemize}[leftmargin=*]
    \item Captures the suboptimality cost due to decentralization, \ie due to communication-minimal distributed coordination.  For example, if the network returned by the algorithm is fully connected, then the algorithm's approximation bound becomes $1/2$. This is near-optimal since the best approximation bound for \cref{eq:intro} is $1-1/e\simeq 0.63$~\cite{Feige:1998:TLN:285055.285059}. 
    \item Holds true given any network topology, including directed and (partially) disconnected networks.  In contrast, the current algorithms, such as the Sequential Greedy algorithm~\cite{fisher1978analysis}, cannot offer suboptimality guarantees when the network is disconnected since they require the agents to be able to relay information about all others.
\end{itemize}

\myParagraph{Numerical Evaluations}
We evaluate \alg in simulated scenarios of area monitoring with multiple cameras (\Cref{sec:experiments}). 
We evaluate the decision speed of and total area covered by \alg with different maximum neighborhood sizes and compare it with the state-of-the-art algorithm \dfs~\cite{konda2022execution}. The results are presented in Fig.~\ref{fig:results}.

\section{Distributed Simultaneous Coordination and Network Design}\label{sec:problem}

We define the problem \textit{Distributed Simultaneous Coordination and   Network Design}. To this end, we use the notation:
\begin{itemize}
    \item $\calV_\calN \triangleq \prod_{i\myin \calN} \,\calV_i$ 
    is the cross product of sets $\{\calV_i\}_{i\myin \calN}$.
    \item $[T]\triangleq\{1,\dots,T\}$ for any positive integer $T$;
    \item $f(\,a\,|\,\calA\,)\triangleq f(\,\calA \cup \{a\}\,)-f(\,\calA\,)$ is the marginal gain of set function $f:2^\calV\mapsto \mathbb{R}$ for adding $a \in \calV$ to $\calA \subseteq\calV$.
    \item $|\calA|$ is the cardinality of a discrete set $\calA$. 
    \item $\calE$ is the set of (directed) communication edges among the agents. $\calE$ is designed in this paper.
\end{itemize}
We also lay down the following framework about  the agents' communication network, and their function $f$.

\myParagraph{Communication network} The communication network $\calG=(\calN, \calE)$ among the agents is \textbf{unspecified} a priori, with the goal in this paper being that the agents must optimize the network themselves to best execute the given task. 

The resulting communication network \textit{can be directed and even disconnected}.  When the network is fully connected (all agents receive information from all others), we call it \textit{fully centralized}. In contrast, when the network is fully disconnected (all agents are isolated, receiving information from no other agent), we call it \textit{fully decentralized}.

\myParagraph{Communication neighborhood}  
When a communication channel exists from agent $j$ to agent $i$, \ie $(j\rightarrow i) \in \calE$, then $i$ can receive, store, and process information from $j$.  The set of all agents that $i$ receives information from is denoted by $\calN_i$. We refer to $\calN_i$ as agent $i$'s \textit{neighborhood}.   

\myParagraph{Communication constraints} Each agent $i$ can receive information from up to $\alpha_i$ other agents due to onboard bandwidth constraints. Thus, it must be $|\calN_i|\leq \alpha_i$. 

Also, we denote by $\calM_i$ the set of agents than have agent $i$ within communication reach ---not all agents may have agent $i$ within communication reach because of distance or obstacles. Therefore, agent $i$ can pick its neighbors by choosing at most $\alpha_i$ agents from $\calM_i$. Evidently, $\calN_i\subseteq\calM_i$.

\begin{definition}[Normalized and Non-Decreasing Submodular Set Function{~\cite{fisher1978analysis}}]\label{def:submodular}
A set function $f:2^\calV\mapsto \mathbb{R}$ is \emph{normalized and non-decreasing submodular} if and only if 
\begin{itemize}
\item (Normalization) $f(\,\emptyset\,)=0$;
\item (Monotonicity) $f(\,\calA\,)\leq f(\,\calB\,)$, $\forall\,\calA\subseteq \calB\subseteq \calV$;
\item (Submodularity) $f(\,s\,|\,\calA\,)\geq f(\,s\,|\,{\mathcal{B}}\,)$, $\forall\,\calA\subseteq {\mathcal{B}}\subseteq\calV$ and $s\in \calV$.
\end{itemize}
\end{definition}

Intuitively, if $f(\,\calA\,)$ captures the number of targets tracked by a set $\calA$ of sensors, then the more sensors are deployed, more or the same targets are covered; this is the non-decreasing property.  Also, the marginal gain of tracked targets caused by deploying a sensor $s$ \emph{drops} when \emph{more} sensors are already deployed; this is the submodularity~property.

\begin{definition}[2nd-order Submodular Set Function{~\cite{crama1989characterization,foldes2005submodularity}}]\label{def:conditioning}
$f:2^\calV\mapsto \mathbb{R}$ is \emph{2nd-order submodular} if and only if 
\begin{equation}\label{eq:conditioning}
    f(s\,|\,\calC) - f(s\,|\,\calA\cup\calC) \geq f(s\,|\,\calB\cup\calC) - f(s\,|\,\calA\cup\calB\cup\calC),
\end{equation}
for any \emph{disjoint} $\calA, \calB, \calC\subseteq \calV$ ($\calA \cap \calB \cap \calC =\emptyset$) and  $s\in\calV$.
\end{definition}

Intuitively, if $f(\,\calA\,)$ captures the number of targets tracked by a set $\calA$ of sensors, then \emph{marginal gain of the marginal gains} drops when more sensors are already deployed.

\begin{problem}[Distributed Simultaneous Coordination and Network Design]
\label{pr:online}
Each agent $i\in\calN$ needs to select a neighborhood $\calN_{i}$ of size at most $\alpha_i$, and an action $a_{i}$ such that the agents jointly solve the optimization problem
\begin{equation}\label{eq:problem}
\hspace{-4mm}\max_{\footnotesize\begin{array}{c}
\calN_{i}\subseteq\mathcal{M}_i \\
         \forall\, i\in \calN
\end{array}}\max_{\footnotesize\begin{array}{c}
a_{i}\in\mathcal{V}_i \\
         \forall\, i\in \calN
\end{array}}f(\,\{a_{i}\}_{i\myin \calN}\,)\;\,\text{s.t.}\,\; |\calN_i|\,\leq \alpha_i,\vspace{-1mm}
\end{equation}
where each agent $i$ selects their action \emph{$a_{i}$ after coordinating actions with its neighbors only, without having access to information about non-neighbors}, and where $f\colon 2^{\calV_{\calN}}\mapsto \mathbb{R}$ is a normalized, non-decreasing submodular, and 2nd-order submodular set function.
\end{problem}

\Cref{pr:online} implies that the network and action optimizations are coupled: when the network is fully decentralized (all agents coordinate with no other), the achieved value of $f$ can be lower compared to the value that can be achieved when the network is instead fully centralized (all agents coordinate with all others). 
For example, consider the target monitoring scenario in \Cref{sec:Intro}: in the fully decentralized setting, all cameras may end up covering the same targets, thus $f$ will equal the number of targets covered by one camera only.  In contrast, in the fully centralized setting, the cameras can coordinate and end up covering different targets, thus maximizing the total number of covered targets $f$.

\begin{remark}[Decision speed vs.~Optimality]\label{rem:resource-awareness}
As demonstrated by the above example, the more centralized a network is, the higher the value of $f$ that can be achieved. 
But a more centralized network can also lead to lower decision speeds due to an explosion of information passing among all the agents since all agents will coordinate with all others.  The goal of this paper is to develop a communication-efficient distributed algorithm that not only requires just a few coordination rounds for convergence; it also needs only short messages to be communicated among agents, thus, accounting for real-world communication delays due to limited communication speeds~\cite{oubbati2019routing}.
For this reason, in particular, \Cref{pr:online} requires each agent to (i) coordinate actions only with its neighbors, and (ii) receive information only about them, instead of also about non-neighbors.  
This is in contrast to standard distributed methods that allow information about the whole network to travel to all agents via multi-hop communication, hence often \underline{not} reducing the amount of information flowing in the network compared to fully centralized coordination, and hence often introducing impractical communication delays~\cite{xu2022resource,wu2021comprehensive}.  
\end{remark}

\vspace{-1mm}\section{Alternating Coordination and Network-Design Algorithm (\alg)} \label{sec:algorithm}

\setlength{\textfloatsep}{3mm}
\begin{algorithm}[t]
	\caption{AlterNAting COordination and Net- work-Design Algorithm (\alg)
	}
	\begin{algorithmic}[1]
		\REQUIRE \!Number of time steps $T$; agent $i$'s neighbor candidate set $\calM_i$; agent $i$'s neighborhood size $\alpha_i$; objective set function $f:2^{\mathcal{V}_\calN} \mapsto \mathbb{R}$. 
		\ENSURE \!Agent $i$'s action $a_{i,\, t}$ and neighbor set $\calN_{i,\, t}$ at each $t\in[T]$.
		\medskip
            \STATE $\calN_{i,\,0}\gets\emptyset, \forall i \in \calN$; 
		\FOR{each time step $t\in [T]$}
   \STATE $a_{i,\,t}\gets\text{\actionsel}([T], \calV_i, f)$;
                \STATE $\calN_{i,\,t}\gets\text{\neighborsel}(a_{i,\,t}, [T], \calM_i, \alpha_i, f)$;
                \STATE \textbf{receive} neighbors' actions $\{a_{j,\, t}\}_{j\myin\calN_{i,\,t}}$ and \\\textbf{update} \actionsel (per lines 6-8) and \neighborsel (per lines 6-11);
		\ENDFOR
	\end{algorithmic}\label{alg:main}
\end{algorithm}

We present the AlterNAting COordination and Network-Design Algorithm (\alg) for \Cref{pr:online}. 
\alg aims to approximate a solution to \Cref{pr:online} by alternating the optimization for action coordination and neighbor selection.  A description of the algorithm is given in {Fig.~\ref{fig:overview}}. 

Both action coordination and neighborhood selection take the form of Multi-Armed Bandit (MAB) problems, therefore, in the following, we first present the MAB problem (\Cref{subsec:prelim}). Then, we will present the algorithms \actionsel (\Cref{subsec:action}) and \neighborsel (\Cref{subsec:neighbor}).

\subsection{Multi-Armed Bandit Problem}\label{subsec:prelim}

The adversarial \textit{Multi-Armed Bandit} (MAB) problem \cite{lattimore2020bandit} involves an agent selecting a sequence of actions to maximize the total reward over a given number of time steps~\cite{lattimore2020bandit}.  The challenge is that, at each time step, the reward associated with each action is unknown to the agent a priori, becoming known only after the action has been selected. To rigorously present the MAB problem, we use the notation:
\begin{itemize}[leftmargin=*]
    \item $\calV$ denotes the available action set;
    \item $v_{t}\in\calV$ denotes the agent's selected action at time $t$;
    \item $r_{v_t,\,t}\in[0,1]$ denotes the reward that the agent receives by selecting action $v_{t}$ at $t$.
\end{itemize}

\begin{problem}[Multi-Armed Bandit~\cite{lattimore2020bandit}]\label{pr:MAB}
Assume an operation horizon of $T$ time steps. At each time step $t\in[T]$, the agent must select an action $v_t$ such that the regret
\begin{equation}\label{eq:MAB}
    \operatorname{MAB-Reg}_T \triangleq\max_{v\myin\mathcal{V}} \;\; \sum_{t=1}^T\;r_{v,\,t} \;\;- \;\;\sum_{t=1}^T\;r_{v_t,\,t},
\end{equation}
is sublinear in $T$, where for \textbf{full-information feedback}, the rewards $r_{v,\,t}\in[0,1]$ for all $ v\in\calV$ become known to the agent \emph{after} $v$ has been executed at each $t$; whereas for \textbf{bandit feedback}, only the reward $r_{v_t,\,t}\in[0,1]$ becomes known to the agent \emph{after} $v$ has been executed.
\end{problem}

\Cref{pr:MAB} asks for $\operatorname{MAB-Reg}_T$ to be sublinear, \ie $\operatorname{MAB-Reg}_T/T\rightarrow 0$ for $T\rightarrow +\infty$, since this implies that the agent asymptotically chooses optimal actions even though the rewards are unknown a priori~\cite{lattimore2020bandit}.

\Cref{pr:MAB} presents two versions of MAB, one with full-information feedback, and one with bandit feedback. The difference between them is that, at each  $t\in[T]$, in full-information feedback the rewards of all $v\in\calV$ are revealed, even though only one action is selected; while in bandit feedback, only the reward of the selected action is revealed. 

\vspace{-1mm}
\subsection{Action Coordination}\label{subsec:action}

\setlength{\textfloatsep}{3mm}
\begin{algorithm}[t]
	\caption{\actionsel
	}
	\begin{algorithmic}[1]
		\REQUIRE \!Number of time steps $T$; agent $i$'s action set $\mathcal{V}_i$; objective set function $f:2^{\mathcal{V}_\calN} \mapsto \mathbb{R}$. 
		\ENSURE \!Agent $i$'s action $a_{i,\, t}$ at each $t\in[T]$.
		\medskip
            \STATE $\eta_1\gets\sqrt{8\log{|\calV_i|}\,/\,{T}}$;
            \STATE $w_{1}\gets\left[w_{1,\,1}, \dots, w_{|\calV_i|,\,1}\right]^\top$ with $w_{v,\,1}=1, \forall a\in \calV_i$;
            \FOR {\text{each time step} $t\in [T]$}
		\STATE \textbf{get} distribution $\distfsf_t\gets{w_t}\,/\,{\|w_t\|_1}$; 
		\STATE \textbf{draw} action $a_{i,\,t}\in\calV_i$ \textbf{from} $\distfsf_t$;
		\STATE \textbf{input} $a_{i,\,t}$ \textbf{to} \neighborsel and \\
        \textbf{receive} neighbors' actions $\{a_{j,\, t}\}_{j\myin\calN_{i,\,t}}$;
		\STATE $r_{a,\, t}\gets f(\,a\,|\, \{a_{j,\, t}\}_{j\myin\calN_{i,\,t}}\,)$ and \\
  \textbf{normalize $r_{a,\, t}$ to} $[0,1]$, $\forall a\in \calV_i$;
            \STATE $w_{a,\,t+1}\gets w_{a,\,t}\exp{(\eta_1 \,r_{a,\,t})}, \forall a\in\calV_i$;		
            \ENDFOR
	\end{algorithmic}\label{alg:action}
\end{algorithm}

We present \actionsel and its performance guarantee.
To this end, we introduce the coordination problem that \actionsel instructs the agents to simultaneously solve and show that it takes the form of \Cref{pr:MAB} with full-information feedback. 
We use the definitions:
\begin{itemize}[leftmargin=*]
    \item $\calA_t\triangleq \{a_{i,\,t}\}_{i\,\in\,\calN}$ is the set of all agents' actions at $t$;
    \item $\solopt\in\arg\max_{a_{i}\myin\mathcal{V}_{i},\, \forall\, i\myin\calN} f(\{a_{i}\}_{i\,\in\, \calN})$ is the optimal actions for agents $\calN$ that solve \cref{eq:intro};
    \item $\calN_{i}^{\star}$ is the optimal neighborhood corresponding to $\{a_{i,\,t}\}_{t\in [T]}$ that solves \cref{eq:neighbor_selection};
    \item $\kappa_f\triangleq 1-\min_{\elem\myin\calV}{[f(\calV)-f(\calV\setminus\{\elem\})]}/{f(\elem)}$ is the curvature of $f$~\cite{conforti1984submodular}.  $\kappa_f$ measures how far~$f$ is from modularity: if $\kappa_f=0$, then  $f(\calV)-f(\calV\setminus\{v\})=f(v)$, $\forall v\in\calV$, \ie $f$ is modular. In~contrast, $\kappa_f=1$ in the extreme case where there exist $v\in\calV$ such that $f(\calV)=f(\calV\setminus\{v\})$, \ie $v$~has no contribution in the presence of $\calV\setminus\{v\}$.  
\end{itemize}

The intuition is that the agents should \emph{select actions simultaneously} such that each agent $i$ selects an action $a_{i,\,t}$ that maximizes the marginal gain $f(\,a\,|\,\{a_{j,\,t}\}_{j\myin\calN_{i,\,t}}\,)$. 
But since the agents select actions simultaneously, 
$\{a_{j,\,t}\}_{j\myin\calN_{i,\,t}}$ become known only after agent $i$ selects $a_{i,\,t}$ and communicates with $\calN_{i,\,t}$, \ie computing $f(\,a\,|\,\{a_{j,\,t}\}_{j\myin\calN_{i,\,t}}\,)$ becomes feasible for all $a\in\calV_i$ only in hindsight. 
To this end, \actionsel instructs each agent $i$ to select actions $\{a_{i,\,t}\}_{t\myin[T]}$ such that the action regret
\begin{equation}\label{eq:action}
    \max_{a\myin\calV_i} \sum_{t=1}^{T} f(\,a\,|\,\{a_{j,\,t}\}_{j\myin\calN_{i,\,t}}\,) - \sum_{t=1}^{T} f(\,a_{i,\,t}\,|\,\{a_{j,\,t}\}_{j\myin\calN_{i,\,t}}\,),
\end{equation}is sublinear in $T$.  Thus, the action coordination problem takes the form of \Cref{pr:MAB} with full-information feedback, where the reward of each action $a\in\calV_i$ is the marginal gain, \ie $r_{a,\,t}\triangleq f(\,a\,|\,\{a_{j,\,t}\}_{j\myin\calN_{i,\,t}}\,)$.

\actionsel implements a  Multiplicative Weights Update (MWU) procedure to converge to an optimal solution to \cref{eq:action} ---the MWU procedure has been introduced to solve \Cref{pr:MAB} with full-information feedback~\cite{cesa2006prediction}. 
In more detail, 
\actionsel starts by initializing a learning rate $\eta_1$ and a weight vector $w_t$ for all available actions $a\in\calV_i$ (\Cref{alg:action}'s lines 1-2). Then, at each time step $t\in[T]$, \actionsel executes in sequence the steps:
\begin{itemize}[leftmargin=*]
    \item Compute probability distribution $p_t$ using $w_{t}$ (lines 3-4);
    \item Select action $a_{i,\,t}\in\calV_i$ by sampling from $p_t$ (line 5);
    \item Send $a_{i,\,t}$ to \neighborsel and receive neighbors' actions $\{a_{j,\, t}\}_{j\myin\calN_{i,\,t}}$ (line 6);
    \item Compute marginal gain of selecting $a\in\calV_i$ as reward $r_{a,\,t}$ associated with each $a\in\calV_i$, and update weight $w_{a,\,t+1}$ for each $a\in\calV_i$ (lines 7-9).\footnote{{The coordination algorithms in~\cite{du2022jacobi,rezazadeh2023distributed,robey2021optimal} instruct the agents to select actions simultaneously at each time step as {\fontsize{7}{7}\selectfont\sf ActionCoordination}, but they lift the coordination problem to the continuous domain and require each agent to know/estimate the gradient of the multilinear extension of $f$, which leads to a decision time at least one order higher than {\fontsize{7}{7}\selectfont\sf ActionCoordination}~\cite{xu2022resource}.}}
\end{itemize}

\begin{proposition}[Approximation Performance]\label{prop:action}
The agents select actions via \actionsel such that:
{
\begin{align}
    &\sum_{t=1}^T f(\,\calA_t\,)\geq\frac{1-\curv_f}{1+\curv_f-\curv_f^2}\sum_{t=1}^T\Bigg[f(\,\solopt\,)\label{eq:action_selection}\\\nonumber 
    &+\sum_{i\myin\calN} \underbrace{[f(a_{i,\,t}) - f(a_{i,\,t}\,|\, \{a_{j,\,t}\}_{j\myin\calN_{i,\,t}})]}_{\smi{a_{i,\,t}}{\calN_{i,\,t}}}\Bigg]-\underbrace{\Tilde{O}\left(|\calN|\sqrt{T}\right)}_{\text{sublinear regret}}. 
\end{align}}where $\tilde{O}(\cdot)$ hides $\log$~terms.
\end{proposition}

\Cref{prop:action} implies that the approximation performance of \actionsel increases for network designs $\{\calN_{i,\,t}\}_{i\myin\calN}$ with higher value $\smi{a_{i,\,t}}{\calN_{i,\,t}}$.  Particularly, the $-\sum_{i \myin \calN}
\smi{a_{i,\,t}}{\calN_{i,\,t}}$ plays the role of $C(\{\calN_{i}\}_{i\myin\calN})$ in Introduction. Intuitively, $\smi{a_{i,\,t}}{\calN_{i,\,t}}$ captures the utility overlap between agent $i$'s action and the actions of its neighbors: for example, when the network is fully disconnected ($\calN_{i,\,t}=\emptyset, \, \forall i\in\calN$), then $\smi{a_{i,\,t}}{\calN_{i,\,t}}=0$.

\begin{definition}[Mutual Information between an Agent and Its Neighbors]\label{def:MI}
At any $t\in [T]$, given an agent $i\in\calN$ with an action $a_{i,\,t}$ and neighbors $\calN_{i,\,t}$, the mutual information between the agent and its neighbors is denoted by\footnote{The quantity in eq.~\eqref{eq:SubMI} extends the definition of \emph{submodular mutual information}~\cite{iyer2021generalized} to the multi-agent setting introduced herein.}
\begin{equation}\label{eq:SubMI}
    \smi{a_{i,\,t}}{\calN_{i,\,t}}\triangleq f(a_{i,\,t}) -f(a_{i,\,t}\,|\,\{a_{j,\,t}\}_{j\myin\calN_{i,\,t}}).
\end{equation}
\end{definition}

\neighborsel will next leverage \Cref{lem:SubMI} to enable the agents to distributively select a network topology that optimizes the approximation bound of \actionsel.

\begin{lemma}[Monotonicity and Submodularity of $I_{f,\,t}$]\label{lem:SubMI}
     Given an $a\in\calV_i$ and a non-decreasing and 2nd-order submodular function $f\colon2^{\calV_\calN}\mapsto \mathbb{R}$, then $\smi{a}{\cdot}$ is non-decreasing and submodular in the second argument. 
\end{lemma}

\subsection{Neighbor Selection}\label{subsec:neighbor}

\setlength{\textfloatsep}{3mm}
\begin{algorithm}[t]
	\caption{\neighborsel
	}
	\begin{algorithmic}[1]
		\REQUIRE \!Number of time steps $T$; agent $i$'s neighbor candidate set $\calM_i\subseteq\calN\setminus\{i\}$; agent $i$'s neighborhood size $\alpha_i$; objective set function $f:2^{\mathcal{V}_\calN} \mapsto \mathbb{R}$.
		\ENSURE \!Agent $i$'s neighbors $\calN_{i,\, t}$ at each $t\in[T]$.
		\medskip
  		\STATE $\eta_2\gets\sqrt{2\log{|\calM_i|}\,/\,{(|\calM_i|T)}}$; $\gamma=\eta_2/2$; 
		\STATE $z_{1}^{(k)}\gets\left[z_{1,\,1}^{(k)}, \dots, z_{\alpha_i,\,1}^{(k)}\right]^\top$ with $z_{j,\,1}^{(k)}=1$, $\forall v\in \calM_i, \forall k\in[\alpha_i]$;    
		\FOR {\text{each time step} $t\in [T]$}
            \STATE \textbf{receive} action $a_{i,\,t}$ \textbf{from} \actionsel;
            \FOR {$k = 1, \dots, \alpha_i$} 
            \STATE \textbf{get} distribution $q_{t}^{(k)}\gets z_t^{(k)}\,/\,{\|z_t^{(k)}\|_1}$; 
            \STATE \textbf{draw} agent $j_{t}^{(k)}\in\calM_i$ \textbf{from} $q_{t}^{(k)}$;
            \STATE \textbf{receive} action $a_{j_{t}^{(k)},\, t}$ \textbf{from} $j_{t}^{(k)}$;
            \STATE $r_{j_{t}^{(k)},\, t}\gets \smi{a_{i,\,t}}{\{a_{j_{t}^{(1)},\,t},\dots,a_{j_{t}^{(k)},\,t}\}} - $\\
            \hspace{1.3cm} $\smi{a_{i,\,t}}{\{a_{j_{t}^{(1)},\,t},\dots,a_{j_{t}^{(k-1)},\,t}\}}$ \\and
            \textbf{normalize $r_{j_{t}^{(k)},\, t}$ to} $[0,1]$;
            \STATE $\Tilde{r}_{j,\,t}^{(k)} \gets 1 - \frac{{\bf 1}(j_{t}^{(k)}\,=\,j)}{q_{j,\,t}^{(k)}\,+\,\gamma}\left(1\,-\,r_{j_{t}^{(k)},\,t}\right)$, $\forall j\in\calM_i$;
            \STATE $z_{j,\,t+1}^{(k)}\gets z_{j,\,t}^{(k)}\exp{(\eta_2 \,\Tilde{r}_{j,\,t}^{(k)})}$, $\forall j\in\calM_i$;					\ENDFOR
                \STATE $\calN_{i,\,t}\gets\{j_{k,\, t}\}_{k\myin [\alpha_i]}$;
		\ENDFOR
	\end{algorithmic}\label{alg:neighbor}
\end{algorithm}

We present \neighborsel. To this end, we introduce the neighbor-selection problem that \neighborsel instructs the agents to simultaneously solve and show that it takes the form of \Cref{pr:MAB} with bandit feedback.

Since \actionsel's suboptimality bound improves when $\smi{a_{i,\,t}}{\calN_{i,\,t}}$ increases, \neighborsel instructs each agent $i$ to select its neighbors by solving the cardinality-constrained maximization problem:
{\begin{equation}\label{eq:neighbor_selection}
    \max_{\calN_{i,\,t}\,\subseteq\,\calM_i,\,|\calN_{i,\,t}|\,\leq\,\alpha_i} \quad \sum_{t=1}^{T} \;\smi{a_{i,\,t}}{\calN_{i,\,t}},
\end{equation}}where $a_{i,\,t}$ is given by \actionsel (Fig.~\ref{fig:overview}). {The problem in \cref{eq:neighbor_selection} is a submodular optimization problem since we prove that $ \smi{a_{i,\,t}}{\calN_i}$ is submodular in $\calN_i$.

But $\smi{a_{i,\,t}}{\calN_{i,\,t}}$ is computable in hindsight only: the $\{a_{j,\,t}\}_{j\myin\calN_{i,\,t}}$ become known only after agent $i$ has selected and communicated with $\calN_{i,\,t}$. Therefore, \cref{eq:neighbor_selection} takes the form of cardinality-constrained bandit submodular maximization~\cite{zhang2019online,matsuoka2021tracking,xu2023bandit}, which is an extension of \Cref{pr:MAB} to the submodular multi-agent setting. 

Solving \cref{eq:neighbor_selection} using algorithms for \Cref{pr:MAB} with bandit feedback will lead  to exponential-running-time algorithms due to an exponentially large $\calV$ per \cref{eq:MAB}~\cite{matsuoka2021tracking}. Therefore, \neighborsel instead extends \cite[Algorithm 2]{matsuoka2021tracking}, which can solve \cref{eq:neighbor_selection} in the full-information setting, to the bandit setting~\cite{xu2023bandit}. Specifically, \neighborsel decomposes \cref{eq:neighbor_selection} to $\alpha_i$ instances of \Cref{pr:MAB} with bandit feedback, and separately solves each of them using the \scenario{EXP3-IX} algorithm~\cite{neu2015explore}, which can handle bandit feedback.

\neighborsel starts by initializing a learning rate $\eta_2$ and $\alpha_i$ weight vectors $z_t^{(k)}, \forall k\in[\alpha_i]$, each determining the $k$-th selection in $\calN_{i,\,t}$ (\Cref{alg:neighbor}'s lines 1-2). Then, at each $t\in[T]$, \neighborsel executes the steps:
\begin{itemize}[leftmargin=*]
    \item Receive action $a_{i,\,t}$ by \actionsel (lines 3-4);
    \item Compute distribution $q_t^{(k)}$ using $z_{t}^{(k)}, \forall k\in [\alpha_i]$ (lines 5-6);
    \item Select agent $j_{t}^{(k)}\in\calM_i$ as neighbor by sampling from $q_t^{(k)}$, and receive its action $a_{j_{t}^{(k)},\, t}, \forall k\in [\alpha_i]$ (lines 7-8);
    \item For each $k\in[\alpha_i]$, compute reward $r_{j_{t}^{(k)},\,t}$ associated with each $j_{t}^{(k)}$, estimate reward $\tilde{r}_{j,\,t}^{(k)}$ for each $j\in\calM_i$,
    and update weight $z_{j,\,t+1}^{(k)}$ for each $j\in\calM_i$ (lines 9-12).
\end{itemize}

\section{Approximation Guarantee}\label{sec:tracking-regret}

We present the suboptimality bound of \alg. 
Thus, \textit{the bound compares \alg with an optimal fully centralized algorithm that maximizes $f$ per \cref{eq:intro}.}
 
\begin{theorem}[Approximation Performance]\label{th:main} \alg instructs over $t\in [T]$ each agent $i\in\calN$ to select actions $\{a_{i,\,t}\}_{t\myin[T]}$ and neighborhoods $\{\calN_{i,\,t}\}_{t\myin[T]}$ that guarantee:
\begin{itemize}[leftmargin=3.5mm]
    \item If the network is \textbf{fully centralized},  \ie $\calN_{i,\,t}\equiv\calN\setminus\{i\}$,
    {\small\begin{equation}\label{eq:thm-1}
        \mathbb{E}\left[f(\,\calA_t\,)\right] \geq \frac{1}{1+\curv_f}\,f(\,\solopt\,)- \underbrace{ \Tilde{O}\left(|\calN|\sqrt{1/T}\right)}_{{\phi(T)}}.
    \end{equation}}
    \item If the network is \textbf{fully decentralized}, \ie $\calN_{i,\,t}\equiv\emptyset$, 
    {\small\begin{equation}\label{eq:thm-2}
        \mathbb{E}\left[f(\,\calA_t\,)\right] \geq \frac{1-\curv_f}{1+\curv_f-\curv_f^2}\,f(\,\solopt\,) -\underbrace{ \Tilde{O}\left(|\calN|\sqrt{1/T}\right)}_{{\chi(T)}}.
    \end{equation}}
    \item If the network is \textbf{anything in between} fully centralized and fully decentralized, \ie $\calN_{i,\,t}\subseteq\calM_i\subseteq\calN\setminus\{i\}$, 
    {\small\begin{align}
    &\hspace{-1mm}\mathbb{E}\left[f(\,\calA_t\,)\right]\geq\frac{1-\curv_f}{1+\curv_f-\curv_f^2}f(\,\solopt\,) \nonumber\\
    &\hspace{-5mm}+\frac{1-\curv_f}{1+\curv_f-\curv_f^2}\frac{1}{\kappa_f}(1-e^{-\kappa_f})\,\mathbb{E}\Bigg[\underbrace{\sum_{i\myin\calN} \smi{a_{i,\,t}}{\calN_{i}^{\star}}}_{I^\star}\Bigg]\nonumber\\\label{eq:thm-4}
    &\hspace{-5mm}\underbrace{- \Tilde{O}\left(\bar{\alpha}|\calN|\sqrt{{|\bar{\calM}|}/{T}}\right) - \log{\left({2}/{\delta}\right)}\Tilde{O}\left(\bar{\alpha}|\calN|\sqrt{{|\bar{\calM}|}/{T}}\right)}_{\psi(T)}, 
    \end{align}}where $\bar{\alpha}\triangleq\max_{i\in\calN}\alpha_i$, and $|\bar{\calM}|\triangleq\max_{i\in\calN}|\bar{\calM_i}|$. 
\end{itemize}
Particularly, each case above holds with probability at least $1-\delta$, for any $\delta \in (0,1)$, and the expectation is due to \alg's internal randomness. $\tilde{O}(\cdot)$ hides $\log$~terms.
\end{theorem}

\Cref{th:main} quantifies both the suboptimality of \alg due to decentralization,  and the convergence speed of \alg. 
Both are affected as follows:
\begin{itemize}[leftmargin=*]
    \item \textit{Effect of online co-design of network topology and agent actions}: $\psi$ in \cref{eq:thm-4} captures the convergence speed of the network and action selection and its impact to the suboptimality bound ---{similarly in \cref{eq:thm-1,eq:thm-2} for $\phi$ and $\chi$}.  Particularly,  $\psi$ vanishes as $T\to\infty$, having no impact on the suboptimality bound anymore, and its vanishing speed captures how fast the agents converge to near-optimal actions and neighborhoods. 
    \item \textit{Effect of resource-minimal distributed communication and action coordination}: After $\psi$ vanishes as $T\to\infty$, the bound in \cref{eq:thm-4} depends on $I^\star$ that captures the suboptimality due to decentralization such that the higher $I^\star$ is the better is \alg's approximation performance.  Particularly, $I^\star$ depends on the neighborhoods of each agent $i$, and  the larger agent $i$'s neighborhood can be, the higher $I^\star$ can be since $\smi{a_{i,\,t}}{\calN_i}$ is non-decreasing in $\calN_i$.
    Then, the better \alg's suboptimality bound can be per \cref{eq:thm-4}. {In contrast, if $\alpha_i=0$ for all agents $i \in \calN$, then $\smi{a_{i,\,t}}{\emptyset}=0$, and as $T\to\infty$, \cref{eq:thm-2} and \cref{eq:thm-4} become the same.}  
    
    Overall, \cref{eq:thm-1,eq:thm-2,eq:thm-4} imply that \alg's global suboptimality bound will improve if the agents can communicate and coordinate over a  more centralized network, with the bound covering the range $[{(1-\curv_f)}/{(1+\curv_f-\curv_f^2)}, {1}/{(1+\curv_f)}]$ as the network covers the spectrum from being fully decentralized (\cref{eq:thm-2}) to fully centralized (\cref{eq:thm-1}).  Importantly, the $1/(1+\curv_f)$ suboptimality bound in the fully centralized case {recovers the bound in~\cite{conforti1984submodular}} and is near-optimal since the best possible bound for in~\eqref{eq:problem} is $1-\kappa_f/e$~\cite{sviridenko2017optimal}.\footnote{{The bounds $1/(1+\curv_f)$ and $1-\kappa_f/e$ become $1/2$ and $1-1/e$ when, in the worst case, $\kappa_f=1$.}}
\end{itemize}
In all, asymptotically (as $T\to\infty$), \textit{\alg enables the agents to individually select near-optimal actions and communication neighborhoods.}

\begin{remark}[On the Tightness of Approximation Bounds]
 The approximation bound in \Cref{th:main} are not  necessarily tight.  Particularly, the bound in \cref{eq:thm-4} does not converge to $1/(1+\kappa_f)$ when the network becomes fully centralized. As part of our future work, we will investigate tight bounds.
\end{remark}

\section{Runtime Analysis}\label{sec:resource-guarantees}
We present the runtime of \alg by analyzing its computation and communication complexity (accounting for message length). 
We use the notation and observations:
\begin{itemize}[leftmargin=*]
    \item $\tau_f$ is the time required for one evaluation of $f$;
    \item $\tau_c$ is the time for {transmitting the information about one action} from an agent directly to another agent;
    \item  $\epsilon$ is \alg's convergence error after $T$ iterations: if the network is fully centralized or fully decentralized, then \alg's convergence error after $T$ iterations is $\epsilon$ only if, per \cref{eq:thm-1,eq:thm-2}, $\phi(T)\leq \epsilon$ or $\chi(T)\leq \epsilon$ , \ie $T\geq |\calN|^2\,/\,\epsilon$.  Similarly, if the network is anything in between, per \cref{eq:thm-4}, only if $\psi(T)\leq \epsilon$, \ie $T\geq |\calM_i|\,|\calN|^2\,/\,\epsilon$.
\end{itemize}

\begin{proposition}[Computational Complexity]\label{prop:computation}
At each $t\in [T]$, \alg requires each agent $i$ to execute $|\calV_i|+2\alpha_i+1$ evaluations of $f$ and $O(|\calV_i|+\alpha_i|\calM_i|)$ additions/multiplications.
\end{proposition}

We prove the proposition here: at each $t\in[T]$, \actionsel  requires $|\calV_i|$ function evaluations  (\Cref{alg:action}'s line 7), along with $O(|\calV_i|)$ additions and multiplications  (\Cref{alg:action}'s lines 4 and 8).  Additionally, \neighborsel requires $2\alpha_i+1$ function evaluations  (\Cref{alg:neighbor}'s line 9), along with $O(\alpha_i|\calM_i|)$ additions and multiplications (\Cref{alg:neighbor}'s lines 6 and 9-11). 

\begin{proposition}[Communication Complexity]\label{prop:communication}
At each $t\myin [T]$, \alg requires~one  communication round where each agent $i$ only transmits its own action to other agents.
\end{proposition}
\Cref{prop:communication} holds true since, at each $t\in[T]$, \alg requires one (multi-channel) communication round where each agent $i$ transmits its $a_{i,\, t}$
(\Cref{alg:main}'s line 11). 

\begin{theorem}[Decision Speed]\label{th:speed}
\alg terminates in $O\left\{[\tau_f\,\max_{i\in \calN}\,(|\calV_i|\,+\,\alpha_i)\,+\,\tau_c\,]\,(\max_{i\in \calN}\,|\calM_i|)\,|\calN|^2\,/\,\epsilon\right\}$.
\end{theorem}

\Cref{th:speed}  holds true by combining  \Cref{prop:computation,prop:communication}, along with the definition of $\epsilon$ above, upon ignoring the time needed for additions and multiplications.

\begin{corollary}[Decision Speed for Sparse Networks]\label{cor:speed}
In sparse networks, where $|\calM_i|\,=o(|\calN|)$, \alg terminates in $O\left\{\,[\tau_f\,\max_{i\in \calN}\,(|\calV_i|\,+\,\alpha_i)\,+\,\tau_c\,]\,|\calN|^2\,/\,\epsilon\,\right\}$ time.
\end{corollary}

\newcommand{\introFigTitleWidth}{0.2cm}
\newcommand{\introFigColWidth}{5.7cm}
\newcommand{\introFigSpacing}{\hspace{-2mm}}
\newcommand{\intoFigNameSpacing}{}
\newcommand{\advFigColWidth}{6cm}

\begin{figure*}[t!]
    \captionsetup{font=footnotesize}
	\begin{center}
	\hspace{-9.8cm}
      \begin{minipage}{\columnwidth}
            \begin{tabular}{p{\introFigColWidth}p{\introFigColWidth}p{\introFigColWidth}}
            \begin{minipage}{\introFigColWidth}%
                  \centering%
                  \includegraphics[width=\columnwidth]{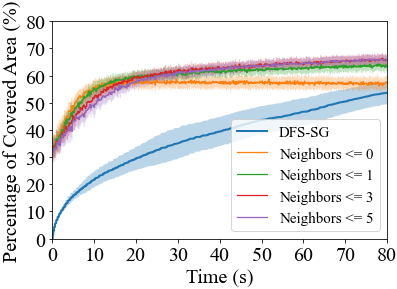} \\
                  \caption*{(a) {$\tau_f = 0.01s, \tau_c = 0.05s$.}
                  }
            \end{minipage}
            &
            \begin{minipage}{\introFigColWidth}
                  \centering%
                  \includegraphics[width=\columnwidth]{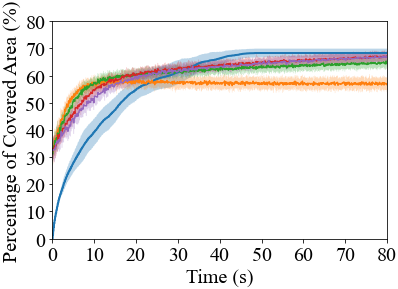} \\ 
                  \caption*{(b) {$\tau_f = 0.01s, \tau_c = 0.01s$.}
                  }
            \end{minipage}
            &
            \begin{minipage}{\introFigColWidth}%
                  \centering%
                  \includegraphics[width=\columnwidth]{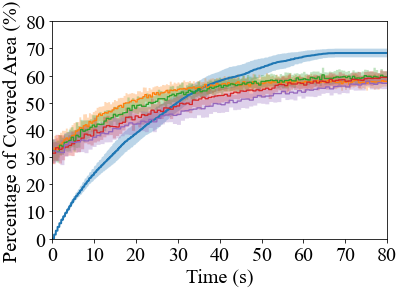} \\
                  \caption*{
                  (c) {$\tau_f = 0.05s, \tau_c = 0.01s$.}
                  }
            \end{minipage}
            \end{tabular}
	\end{minipage} 
	\caption{\textbf{Area Monitoring with Multiple Cameras: {\smaller \sf Anaconda} vs. {\smaller \sf DFS-SG}.}   
	The cameras select the locations of their FOVs either per {\smaller \sf Anaconda} with different maximum neighborhood sizes ranging among $\{0,1,3,5\}$, or per the {\smaller \sf DFS-SG}. (a)-(c) are averaged over $30$ Monte-Carlo trials. 
 From (a) to (b) to (c), the time $\tau_c$ of communicating an action decreases compared to the time $\tau_f$ of completing a function evaluation, with $\tau_c/\tau_f=\{5, 1, 0.2\}$.
	}\label{fig:results}
	\vspace{-7mm}
	\end{center}
\end{figure*}

\begin{remark}[\alg vs.~Sequential Greedy~\cite{fisher1978analysis}]\label{rem:vsSG}
In sparse networks, \alg can be one order faster than the Sequential Greedy algorithm. Particularly, in Appendix II, we prove that for directed networks, which are the focus in this paper, the Sequential Greedy requires $O(\,\tau_f\,\sum_{i\in \calN}\,|\calV_i|\,+\,\tau_c\,|\calN|^3)$ time to terminate since the agents must sequentially choose actions where the $i$-th agent in the sequence needs to evaluate $f$ for $|\calV_i|$ times and to transmit to the $(i+1)$-th agent all the actions selected so far, \ie $\{a_{s,\,t}\}_{s\myin[i]}$ ---{the proof follows the steps of~\cite[Proposition~2]{konda2022execution} accounting for the size of each inter-agent communication message}.
In contrast, \alg requires $O\left\{\,[\tau_f\,\max_{i\myin \calN}\,(|\calV_i|\,+\,\alpha_i)\,+\,\tau_c\,]\,|\calN|^2\,/\,\epsilon\,\right\}$ time to terminate. For example, if $\tau_f=\tau_c=\tau$ and $|\calV_i|\,=v, \forall i\myin\calN$, with $\tau$ and $v$ being constant, then Sequential Greedy takes $O(|\calN|^3)$ time while \alg takes $O(|\calN|^2/\epsilon)$ time.\footnote{Even when more agents join the network and, thus, $|\calN|$ increases, $|\calV_i|$, $\tau_c$, and $\tau_f$ can remain constant since, respectively, $|\calV_i|$ is intrinsic to the agent $i$, $\tau_c$ is intrinsic to the communication channel (assuming all agents can establish channels with the same speed $\tau_c$), and $\tau_f$ is independent of $|\calN|$ since each agent can communicate with at most $\alpha_i$ other agents independently of the number of existing agents.}
\end{remark}


\section{Numerical Evaluation in Sensor Scheduling for Area Monitoring}\label{sec:experiments}

We evaluate \alg in a simulated scenario of 2D area monitoring. The results are summarized in Fig.~\ref{fig:results}.  
We next elaborate on the simulated area monitoring setup, the compared algorithms, and the used performance metrics.

\myParagraph{Area Monitoring Setup}
The setup is as follows: 
\begin{itemize}[leftmargin=*]
    \item \textit{Environment:} The environment is a static $100\times 100$ map. 
    \item \textit{Agents:} There exist $60$ downward-facing cameras.
    Each camera $i\in\calN$ is located at $x_{i}\in [0,100]^2$ and can point its limited field of view (FOV) to different directions. 
    \item \textit{Actions:} Each camera $i$ has a circular FOV of radius $r=7$ and can point it to 8 cardinal directions. 
    Particularly, at each time step $t$, camera $i$ can locate the center of its FOV at $a_{i,\,t}\in\calV_i$ where $\calV_i\triangleq x_{i}+[r\cos{\theta_t}, r\sin{\theta_t}]$ and $\theta_t$ is one of the 8 cardinal directions. 
    {Each camera $i$ is unaware of $\calV_j, j\in\calN\setminus\{i\}$. Thus, the cameras have to communicate to know about one another's action information.} 
    \item \textit{Communication Network:} The emergent communication network $\calG_t$ can be directed and time-varying. At each time step $t$, each camera $i$ first finds its neighbor candidate set $\calM_i\triangleq\{j\}_{\|x_j-x_i\|\leq c_i,\, j\myin\calN\setminus\{i\}}$, where $c_i$ is $i$'s communication range. Then, it uses \neighborsel to select neighborhood $\calN_{i,\,t}$ from $\calM_i$.  Once $\calN_{i,\,t}$ is determined by all cameras $i\in\calN$, $\calG_t$ is defined. 
    \item \textit{Objective Function:}  $f(\{a_{i,\,t}\}_{i\myin \calN})$ is the total area covered by the cameras $\calN$ when they select $\{a_{i,\,t}\}_{i\myin \calN}$ as the centers of their FOVs. $f$ is proved to be submodular~\cite{corah2018distributed}.
\end{itemize}

\myParagraph{Compared Algorithms}
We evaluate \alg against the state-of-the-art algorithm \dfs~\cite{konda2022execution} across 30 Monte-Carlo scenarios where at each trial (i) each camera location $x_i$ is uniformly sampled from $[0,100]^2$, $\forall i\in\calN$, and (ii) the communication ranges $c_i$, $\forall i\in\calN$ are uniformly sampled from $[15,20]$.
We repeat the 30 trials over three sets of possible values of $\tau_f$ and $\tau_c$, $(\tau_f, \tau_c)=(0.01s,$ $ 0.05s), (0.01s, 0.01s), \text{or } (0.05s, 0.01s)$, that
{capture scenarios with different computational and communication loads.} 

In more detail, we evaluate the following algorithms:
\paragraph{\alg with different neighborhood sizes} To evaluate the impact of decentralization on the trade-off between decision speed and total area coverage, at each Monte-Carlo trial, we ran multiple \alg varying the maximum neighborhood size  $\alpha_i\leq n_{max}$, where $n_{max}=\{0,1,3,5\}$. 

\paragraph{\dfs~\emph{\cite{konda2022execution}}}
We also compare \alg with the state-of-the-art algorithm \dfs. \dfs requires a given connected communication network to run. To this end, at each trial, we first sample the same communication ranges $c_i$ from $[15, 20]$ and construct $\calM_i, \forall i\in\calN$ for both \alg and \dfs. Then, while \alg actively selects neighbors $\calN_{i,\,t}\subseteq\calM_i, \forall i\in\calN$, to enable scalability, \dfs will directly take $\calN_{i,\,t}\equiv\calM_i, \forall t$. We set the range $c_i$ to be not too small to ensure the communication network for \dfs is always connected.

\myParagraph{Performance Metrics} We evaluate the performance of the algorithms in terms of their (i) decision speed ---time to convergence--- and (ii) achieved objective value.

\myParagraph{System Specifications} We ran all simulations using Python 3.11.7 on a Windows PC with the Intel Core i9-14900KF CPU @ 3.20 GHz and 64 GB RAM. 

 \textbf{Code.} 
{The code is available at \href{https://github.com/UM-iRaL/Self-configurable-network}{\blue{https://github.com/UM-iRaL/Self-configurable-network}}.}

\myParagraph{Results} 
The simulation results are presented in Fig.~\ref{fig:results}. From Fig.~\ref{fig:results}, we observe the following:
\begin{itemize}[leftmargin=*]
    \item \textit{Trade-off of centralization and decentralization:} As the neighborhood size $n_{max}$ increases, \ie the coordination becomes more centralized, the convergence speed of \alg decreases, while the total covered area upon convergence increases. These observations agree with the proven theory: (i) per \Cref{prop:computation}, \neighborsel runs faster when each agent can select fewer neighbors; and (ii) per \Cref{th:main}, as the cameras' neighborhood sizes increase, the approximation performance of \alg increases. 
    \item \textit{\alg vs.~\dfs:} Upon convergence, both algorithms cover a comparable total area. As expected, \alg converges faster when the communication cost to the decision speed is no less than the computational cost ($\tau_c\geq \tau_f$).   In more detail, we observe the following: 
    \begin{itemize}
        \item \textit{Total covered area}: \alg starts with a non-zero covered area (30\% of the total area), whereas \dfs starts from near-zero covered area.  The reason is that \alg instructs all cameras to execute an action from the start of time, whereas \dfs instructs the cameras to execute actions sequentially. Upon convergence, the two algorithms cover a comparable total area. 
        \item  \textit{Convergence speed}: When the communication cost to the decision speed is no less than the computation cost ($\tau_c\geq \tau_f$), then Anaconda converges faster.  For example, for $\tau_f=0.01s$ and $\tau_c=0.05s$ (Fig.~\ref{fig:results}(a)), \alg converges within $10$s to $20$s, whereas \dfs achieves comparable performance at $80$s.  \alg converges slower when $\tau_f=0.05s$ and $\tau_c=0.01s$ (Fig.~\ref{fig:results}(c)) since \alg requires more computations per  step.  But still, \alg covers a comparable total area to \dfs across all time steps in Fig.~\ref{fig:results}(c). 
    \end{itemize}
\end{itemize}

\section{Conclusion} \label{sec:con}
We introduced a rigorous approach, \alg, that enables multi-agent networks to self-configure their communication topology to balance the trade-off between decision speed and approximation performance during multi-agent coordination.  
Compared to the state of the art, \alg is an anytime self-configuration algorithm that quantifies its suboptimality guarantee for any type of network, from fully disconnected to fully centralized, and that, for sparse networks, is one order faster.
We demonstrated \alg in simulated scenarios of area monitoring with multiple cameras. 

\myParagraph{Future work} We will extend this work to reduce the number of function evaluations needed by \alg,  and investigate tighter suboptimality bounds.


\bibliographystyle{IEEEtran}
\bibliography{references}



\section{Proofs of \Cref{prop:action} 
 and \Cref{th:main}}\label{app:main}
We first present the proof of \Cref{lem:SubMI} and suboptimality bound of \neighborsel, then present proofs of \Cref{prop:action} and \Cref{th:main}.

\subsection{Proof of \Cref{lem:SubMI}}\label{subsec:app-SubMI}

Consider $\smi{a}{\calJ}= f(a) - f(a\,|\,\{a_{j}\}_{j\myin\calJ})$,  where $\calJ\subseteq\calM_i\subseteq\calN\setminus\{i\}$, $a\in\calV_i$ is fixed, and $f\colon2^{\calV_\calN}\mapsto \mathbb{R}$ is non-decreasing and 2nd-order submodular. Also, with a slight abuse of notation, denote $f(a\,|\,\{a_{j}\}_{j\myin\calJ})$ by $f(a\,|\,\calJ)$. 

To prove the monotonicity of $\smi{a}{\cdot}$, consider $\calA_1, \calA_2\subseteq\calM_i$ that are disjoint.  Then,
$\smi{a}{\calA_1\cup\calA_2} - \smi{a}{\calA_1} = -f(a\,|\,\calA_1\cup\calA_2) + f(a\,|\,\calA_1)\geq 0$ since $f$ is submodular.  Thus, $\smi{a}{\cdot}$ is non-decreasing. 

To prove the submodularity of $\smi{a}{\cdot}$, consider $\calA, \calB_1, \calB_2\subseteq\calV$, where $\calB_1$ and $\calB_2$ are disjoint, then:
{
\begin{align}
   & \hspace{-1.5mm}\smi{a}{\calA\,|\,\calB_1} - \smi{a}{\calA\,|\,\calB_1\cup\calB_2}\nonumber\\
    & \hspace{-2mm}=\smi{a}{\calA\cup\calB_1} - \smi{a}{\calB_1}\nonumber\\
    & \hspace{-2mm}\hspace{1.5cm}- \smi{a}{\calA\cup\calB_1\cup\calB_2} + \smi{a}{\calB_1\cup\calB_2}\nonumber\\
    & \hspace{-2mm}=-f(a\,|\,\calA\cup\calB_1) + f(a\,|\,\calB_1)\nonumber\\
    & \hspace{-2mm}\hspace{1.5cm} + f(a\,|\,\calA\cup\calB_1\cup\calB_2) - f(a\,|\,\calB_1\cup\calB_2)\geq 0,
\end{align}}where the inequality holds since $f$ is 2nd-order submodular (\Cref{def:conditioning}). Therefore, $\smi{a}{\cdot}$ is submodular. 

\subsection{Approximation Performance of \neighborsel}\label{subsec:app-neighbor}
Per \Cref{prop:action}, the approximation performance of \neighborsel can be improved by maximizing $\smi{a}{\cdot}$ online with respect to its second argument subject to a  cardinality constraint and with bandit feedback. 
The suboptimality of \neighborsel is captured by the following definition. 

\begin{definition}[$\kappa_f^{-1}(1-e^{-\kappa_f})$-Approximate Static Regret of Network Design]\label{def:NReg}
    Consider a set of agents $\calJ_t\subseteq\calM$ is selected as neighbors by an agent when its selected action $a_t$ is given a priori, $|\calJ_t|\leq\alpha$, $\forall t\in [T]$. The $\kappa_f^{-1}(1-e^{-\kappa_f})$-approximate static regret of $\{\calJ_t\}_{t\myin[T]}$ is given by:
    \begin{align}\label{eq:regret_1}
        &\operatorname{N-Reg}_{\{a_t\}_{t\in [T]}}^{(\kappa_f^{-1}(1-e^{-\kappa_f}))}(\,\{\calJ_t\}_{t\myin[T]}\,) \\
        &\triangleq \kappa_f^{-1}(1-e^{-\kappa_f})\max_{\calJ\subseteq\calM,\, |\calJ|\leq\alpha}\sum_{t=1}^{T} \smi{a_t}{\calJ} \nonumber\\\label{eq:regret_2}
        &\quad- \sum_{t=1}^{T}\smi{a_t}{\calJ_t}.
    \end{align}
\end{definition}

Eq.~\eqref{eq:regret_1} evaluates $\{\calJ_t\}_{t\myin[T]}$'s suboptimality against the optimal neighborhood that would have been selected if $\smi{a_t}{\cdot}$ had been known a priori $\forall t\in [T]$.  The optimal total value in \cref{eq:regret_1} is discounted by $\kappa_f^{-1}(1-e^{-\kappa_f})$ 
since \Cref{pr:online} is NP-hard to solve with an approximation factor more than $\kappa_f^{-1}(1-e^{-\kappa_f})$ even when $\smi{a_t}{\cdot}$ is known a priori~\cite{conforti1984submodular}. 

We next bound \cref{eq:regret_1}:

{
\begin{align}
    &\hspace{-1mm}\mathbb{E}\left[\NReg^{(\kappa_f^{-1}(1-e^{-\kappa_f}))}(\,\{\calJ_t\}_{t\myin [T]}\,)\right] \nonumber    \\\label{aux2:1}
    &\hspace{-2mm}= \mathbb{E}\Bigg[\kappa_f^{-1}(1-e^{-\kappa_f})\max_{\calJ\subseteq\calM,\, |\calJ|\leq\alpha}\sum_{t=1}^{T} \smi{a_t}{\calJ} \nonumber\\
    & \hspace{4.7cm}- \sum_{t=1}^{T}\smi{a_t}{\calJ_t}\Bigg]\\\label{aux2:2}
    &\hspace{-2mm}\leq \kappa_f^{-1}(1-e^{-\kappa_f})\sum_{t=1}^{T} \mathbb{E} \left[ - \alpha\, r_{j_{k,\,t},\, t}+\sum_{k=1}^{\alpha}r_{j_{k,\,t}^\opt,\, t}\right]\\\label{aux2:3}
    &\hspace{-2mm}= \kappa_f^{-1}(1-e^{-\kappa_f})\sum_{k=1}^{\alpha}\mathbb{E} \left[\sum_{t=1}^{T} \left(r_{j_{k,\,t}^\opt,\, t} - r_{j_{k,\,t},\, t}^{\top}\, q_{k,\,t}\right)\right]\\
    \label{aux2:5}
    &\hspace{-2mm}\leq\Tilde{O}\left(\alpha\sqrt{|\calM|T}\right) + \log{\left({2}/{\delta}\right)}\Tilde{O}\left(\alpha\sqrt{|\calM|T}\right),
\end{align}}with probability at least $1-\delta$, where \cref{aux2:1} follows from \cref{eq:regret_1}, \cref{aux2:2} follows from \cite[Theorem 3]{matsuoka2021tracking}, \cref{aux2:3} follows from the linearity of expectation, and \cref{aux2:5} follows by applying \cite[Theorem 1]{neu2015explore}, which provides the suboptimality bound of \scenario{EXP3-IX} that is used in \neighborsel.   

\subsection{Proofs of \Cref{prop:action} and \Cref{th:main}}\label{subsec:app-main}
Before proving \Cref{prop:action} and \Cref{th:main}, we first present the following suboptimality bound for solving \cref{eq:action} using MWU. Denoting the quantity in \cref{eq:action} as $\operatorname{A-Reg}_T(\,\{a_{i,\,t}\}_{t\myin [T]}\,)$, we bound it per~\cite[Theorem 2.2]{cesa2006prediction}:
    {
    \begin{align}
        \AReg(\,\{a_{i,\,t}\}&_{t\myin [T]}\,) \triangleq \sum_{t=1}^{T} \left(r_{a_{i}^\opt,\,t} - r_{a_{i,\,t},\,t}\right)\nonumber\\\label{aux22:8}
        &\leq \sqrt{(T\log{|\calV_i|})/2} = \Tilde{O}\left(\sqrt{T}\right).
    \end{align}}

We now prove the main result, using the preceding preliminary results: 
{\small\begin{align}
    &\sum_{t=1}^{T}f(\calA^\opt)\nonumber\\
    &\leq\kappa_f \sum_{t=1}^{T} f(\calA_t)\nonumber + (1-\kappa_f) \sum_{t=1}^{T} \sum_{a_{i,t}\in\calA_t\cap\calA^\opt} f(a_{i,t}\,|\,\calA_{[i-1],t})\nonumber \\
    & \quad+ \sum_{t=1}^{T} \sum_{a_{i}^\opt\in\calA^\opt\setminus\calA_t} f(a_{i}^\opt\,|\,\calA_t) \label{aux22:1}\\
    &\leq\kappa_f \sum_{t=1}^{T} f(\calA_t)\nonumber + (1-\kappa_f) \sum_{t=1}^{T} \sum_{a_{i,t}\in\calA_t\cap\calA^\opt} f(a_{i,t}\,|\,\calA_{[i-1],t})\nonumber \\
    & \quad+ \sum_{t=1}^{T} \sum_{a_{i}^\opt\in\calA^\opt\setminus\calA_t} f(a_{i}^\opt\,|\,\{a_{j,t}\}_{j\in\calN_{i,t}}) \label{aux22:2}\\
    &\leq\kappa_f \sum_{t=1}^{T} f(\calA_t)\nonumber + \sum_{t=1}^{T} \sum_{a_{i,t}\in\calA_t\cap\calA^\opt} f(a_{i,t}\,|\,\{a_{j,t}\}_{j\in\calN_{i,t}})\nonumber \\
    & \quad+ \sum_{t=1}^{T} \sum_{i\,\colon\, a_{i}^\opt\neq a_{i,t}} \left[f(a_{i}^\opt\,|\,\{a_{j,t}\}_{j\in\calN_{i,t}}) - f(a_{i,t}\,|\,\{a_{j,t}\}_{j\in\calN_{i,t}})\right] \nonumber \\
    & \quad+ \sum_{t=1}^{T} \sum_{i\,\colon\, a_{i}^\opt\neq a_{i,t}} f(a_{i,t}\,|\,\{a_{j,t}\}_{j\in\calN_{i,t}}) \label{aux22:3}\\
    &\leq\kappa_f \sum_{t=1}^{T} f(\calA_t) + \sum_{t=1}^{T} \sum_{i\in\calN} f(a_{i,t}\,|\,\{a_{j,t}\}_{j\in\calN_{i,t}}) \label{aux22:4}\\
    &\quad+ \sum_{t=1}^{T} \sum_{i\in\calN} \left[f(a_{i}^\opt\,|\,\{a_{j,t}\}_{j\in\calN_{i,t}}) - f(a_{i,t}\,|\,\{a_{j,t}\}_{j\in\calN_{i,t}})\right] \nonumber\\
    &=\kappa_f \sum_{t=1}^{T} f(\calA_t)\nonumber + \sum_{t=1}^{T} \sum_{i\in\calN} \left(r_{a_{i}^\opt,t} - r_{a_{i,t},t}\right) + \sum_{t=1}^{T} \sum_{i\in\calN} f(a_{i,t})\nonumber \\
    & \quad- \sum_{t=1}^{T} \sum_{i\in\calN} \left[f(a_{i,t}) - f(a_{i,t}\,|\,\{a_{j,t}\}_{j\in\calN_{i,t}})\right] \label{aux22:5}\\
    &\leq\kappa_f \sum_{t=1}^{T} f(\calA_t)\nonumber + \sum_{i\in\calN} \AReg(\{a_{i,t}\}_{t\in [T]}) + \frac{1}{1-\kappa_f} \sum_{t=1}^{T} f(\calA_t)\nonumber \\
    & \quad- \sum_{t=1}^{T} \sum_{i\in\calN} \smi{a_{i,t}}{\calN_{i,t}} \label{aux22:6}\\
    &=\frac{1+\kappa_f-\kappa_f^2}{1-\kappa_f} \sum_{t=1}^{T} f(\calA_t)\nonumber + \sum_{i\in\calN} \AReg(\{a_{i,t}\}_{t\in [T]}) \nonumber \\
    & \quad - \kappa_f^{-1}(1-e^{-\kappa_f}) \sum_{t=1}^{T} \smi{a_{i,t}}{\calN_{i}^{\star}}  \nonumber \\
    &\quad+\sum_{i\in\calN}\operatorname{N-Reg}_{\{a_{i,t}\}_{t\in [T]}}^{(\kappa_f^{-1}(1-e^{-\kappa_f}))}(\{\calN_{i,t}\}_{t\in [T]}), \label{aux22:7}
\end{align}}where \cref{aux22:1} holds from \cite[Lemma 2.1]{conforti1984submodular}; \cref{aux22:2,aux22:4} hold from $f$ being submodular; \cref{aux22:3} holds since $1-\kappa_f \leq \frac{f(a_{i,t}\,|\,\{a_{j,t}\}_{j\in\calN\setminus\{i\},t})}{f(a_{i,t})} \leq \frac{f(a_{i,t}\,|\,\{a_{j,t}\}_{j\in\calN_{i,t}})}{f(a_{i,t}\,|\,\calA_{[i-1],t})}$ per the definition of $\kappa_f$; 
\cref{aux22:6} holds from \cite[Lemma 2.1]{iyer2013curvature}; and \cref{aux22:7} holds from \cref{eq:regret_2}.

The proof of \Cref{prop:action} follows from \cref{aux22:6}. We continue for \Cref{th:main} by combining \cref{aux2:5,aux22:7,aux22:8}:
{\small
\begin{align}
    &\frac{1+\kappa_f-\kappa_f^2}{1-\curv_f}\sum_{t=1}^{T} f(\calA_t) \geq \sum_{t=1}^{T} f(\solopt)\nonumber\\
    & + \frac{1}{\kappa_f}(1-e^{-\kappa_f})\sum_{t=1}^{T}\sum_{i\in\calN} \smi{a_{i,\,t}}{\calN_{i}^{\star}} - \Tilde{O}\left(|\calN|\sqrt{T}\right) \nonumber\\\label{aux22:10}
    & - \Tilde{O}\left(\bar{\alpha}|\calN|\sqrt{|\bar{\calM}|T}\right) - \log{\left(\frac{2}{\delta}\right)}O\left(\bar{\alpha}|\calN|\sqrt{|\bar{\calM}|T}\right),
\end{align}}with probability at least $1-\delta$. Therefore, \cref{aux22:10} becomes:
{\small
\begin{align}
    &\mathbb{E}\left[f(\calA_t)\right] = \frac{1}{T}\sum_{t=1}^{T} f(\calA_t) \nonumber\\
    &\geq \frac{1-\curv_f}{1+\kappa_f-\kappa_f^2}\frac{1}{T}\sum_{t=1}^{T} f(\solopt)\\\nonumber
    &\quad+ \frac{1-\curv_f}{1+\kappa_f-\kappa_f^2}\frac{1}{\kappa_f}(1-e^{-\kappa_f})\frac{1}{T}\sum_{t=1}^{T}\sum_{i\in\calN} \smi{a_{i,\,t}}{\calN_{i}^{\star}} \\\nonumber
    &\quad- \Tilde{O}\left(\bar{\alpha}|\calN|\sqrt{|\bar{\calM}|/{T}}\right) - \log{\left({2}/{\delta}\right)}O\left(\bar{\alpha}|\calN|\sqrt{{|\bar{\calM}|}/{T}}\right)\\
    &= \frac{1-\curv_f}{1+\kappa_f-\kappa_f^2} f(\solopt) \\\nonumber
    &\quad+ \frac{1-\curv_f}{1+\kappa_f-\kappa_f^2}\frac{1}{\kappa_f}(1-e^{-\kappa_f})\sum_{i\in\calN} \mathbb{E}\left[\smi{a_{i,t}}{\calN_{i}^{\star}}\right] \\\nonumber
    &\quad - \Tilde{O}\left(\bar{\alpha}|\calN|\sqrt{{|\bar{\calM}|}/{T}}\right) - \log{\left({2}/{\delta}\right)}O\left(\bar{\alpha}|\calN|\sqrt{{|\bar{\calM}|}/{T}}\right),
\end{align}}with probability at least $1-\delta$, and, thus, \cref{eq:thm-4} is proved. 

To prove \cref{eq:thm-1}, where $\calN_{i,t}\equiv\calM_i=\calN\setminus\{i\}$: 
{
\begin{align}
    &\kappa_f\,\mathbb{E}\left[f(\calA_t)\right] \geq f(\solopt)\nonumber \\
    &- \sum_{i\in\calN} \mathbb{E}\left[f(a_{i,t}\,|\,\calA_{t}\setminus\{a_{i,t}\})\right]- \Tilde{O}\left(|\calN|\sqrt{1/T}\right)\label{aux222:1}\\
    &\geq f(\solopt) - \mathbb{E}\left[f(\calA_t)\right] - \Tilde{O}\left(|\calN|\sqrt{1/T}\right),\label{aux222:2}
\end{align}}with probability at least $1-\delta$, where \cref{aux222:1} holds from \cref{aux22:4}, and \cref{aux222:2} holds from~\cite[Eq.~(15)]{tzoumas2017resilient}. Thereby, 
\begin{equation}
    \mathbb{E}\left[f(\calA_t)\right] \geq \frac{1}{1+\kappa_f} f(\solopt) - \Tilde{O}\left(|\calN|\sqrt{1/T}\right),
\end{equation}with probability at least $1-\delta$. 

Finally, to prove \cref{eq:thm-2}, where $\calN_{i,t}\equiv\emptyset$, per \cref{aux22:6}, with probability at least $1-\delta$,
\begin{equation}
    \hspace{-1mm}\mathbb{E}\left[f(\calA_t)\right] \geq \frac{1-\curv_f}{1+\kappa_f-\kappa_f^2} f(\solopt) - \Tilde{O}\left(|\calN|\sqrt{1/T}\right).
\end{equation}

\section{Worst-Case Decision Time of Sequential Greedy in Directed Networks}

We prove that \dfs has a $O(\tau_c\,|\calN|^3)$ worst-case communication time on a strongly connected directed graph. The proof extends \cite[Section III-D]{konda2022execution} by taking also the size of each inter-agent communication message into consideration since the larger the size the more time it will take for the message to be transmitted. We use the notations:
\begin{itemize}[leftmargin=3.5mm]
    \item $\calG_{dir}=\{\calN, \calE_{dir}\}$ is a strongly connected directed graph;
    \item $\pi\colon\{1,\dots,|\calN|\}\mapsto\{1,\dots,|\calN|\}$ denotes the order of action selection for agent $i\in\calN$ given by the DFS approach in \cite{konda2022execution};
    \item $d(i,j)$ denotes the length of the shortest path from agent $i$ to $j$ on $\calG_{dir}$.
\end{itemize}
Suppose $p=(v_1,\dots,v_l)$ is the longest path of $\calG_{dir}$, where
$l = |p|$. If $l=|\calN|$, then $p$ is a spanning walk on $\calG_{dir}$ with $\pi(i)=i, \forall i=\{1,\dots,|\calN|\}$ and 
\begin{align}
    &\max_{\calG_{dir}}T_{min}(\calG_{dir}) = \sum_{i\,=\,1}^{|\calN|-1} i\tau_c \times d(i,i+1) \nonumber\\
    &=\sum_{i\,=\,1}^{|\calN|-1} i\tau_c\times 1 =\tau_c\,|\calN|\,(|\calN|-1)/2 \leq O(\tau_c\,|\calN|^3).
\end{align}
Otherwise, the worst-case $\calG_{dir}$ should have $v_l$ being the first vertex of $p$ that is adjacent to a vertex $\bar{v}\in p$,
Then we have
\begin{align}
    &\max_{\calG_{dir}}T_{min}(\calG_{dir}) \nonumber\\
    &= \sum_{i\,=\,1}^{l-1} i\tau_c \times d(i,i+1) + \sum_{i\,=\,l}^{|\calN|-1} i\tau_c \times d(i,i+1) \nonumber\\\label{aux31:1}
    &\leq \sum_{i\,=\,1}^{l-1} i\tau_c\times 1 + \sum_{i\,=\,l}^{|\calN|-1} i\tau_c \times (l-1) \\\label{aux31:2}
    &=\frac{1}{2}\tau_c\,[l(l-1) + (l-1)(|\calN|+l-1)(|\calN|-l)] \\\nonumber
    &= O(\tau_c\,|\calN|^3), 
\end{align}where \cref{aux31:1} holds since no path in $\calG_{dir}$ is longer than $p$, and the maximum of \cref{aux31:2} is taken when $l = \lceil{\sqrt{3|\calN|^2-3|\calN|+3}/3}\rceil$. 
In all, \dfs has a $O(\tau_c\,|\calN|^3)$ worst-case communication time.

\end{document}